\documentstyle[12pt]{article} 
\renewcommand{\theequation}{\arabic{section}.\arabic{equation}}
\begin{document}
\vspace*{-2cm}
\begin{flushright}La Plata-Th 96/10
\end{flushright}
\vspace{1.cm}
\begin{center}
{\Large {\bf Field-theoretical approach to non-local interactions: 1d electrons and
fermionic impurities}}\\
\vspace{1.cm}

{C.M.Na\'on, M.C.von Reichenbach and M.L.Trobo 
\footnote{E-mail adress: naon@venus.fisica.unlp.edu.ar}\\
{\normalsize \it Depto. de F\'\i sica.  Universidad
Nacional de La Plata.}\\  
{\normalsize \it CC 67, 1900 La Plata, Argentina.}\\
{\normalsize \it Consejo Nacional de Investigaciones
Cient\'\i ficas y T\'ecnicas, Argentina.}\\}
\vspace{.5cm}
\large{\it August 1996}
\end{center}

\vspace{2.cm}
\begin{abstract}
We apply a recently proposed path-integral approach to non-local bosonization 
to a Thirring-like system modeling non-relativistic massless particles 
interacting with localized fermionic impurities. We consider forward 
scattering processes described by symmetric potentials including interactions 
between charge, current, spin and spin-current densities. In the general 
(spin-flipping) problem we obtain an effective action for the collective modes 
of the model at T = 0, containing WZW-type terms. When spin-flipping processes
are disregarded the structure of the action is considerably simplified, 
allowing us to derive exact expressions for the dispersion relations of 
collective modes and two point fermionic correlation functions as functionals 
of the potentials. Finally, as an example, we compute the momentum 
distribution for the case in which electrons and impurities are coupled 
through spin and spin-current densities only. The formulae we get suggest that 
our formalism could be useful in order to seek for a mechanism able to restore 
Fermi liquid behavior.

\end{abstract}
\pagenumbering{arabic}

\newpage

\section{Introduction}

In the last years there has been a renewed interest in the study of 
low-dimensional field theories. This is due, in part, to crucial advances 
in the field of nanofabrication which have allowed to build ultranarrow 
semiconductor structures \cite{PT} in which the motion of the electrons is 
confined to one dimension \cite{Voit}. One of the main tools for the 
theoretical understanding of the one-dimensional (1d) electron system is the 
Tomonaga-Luttinger (TL) model \cite{To},\cite{ML}, which can be considered as 
the paradigm of Luttinger Liquid (LL) behavior \cite{H},\cite{Voit}. This 
model describes a non-relativistic gas of massless particles (the electrons) 
with linear free dispersion relation and two-body, forward-scattering 
interactions.

In a recent work \cite{NRT}, we have presented a Thirring-like model 
with fermionic currents coupled by general (symmetric) bilocal potentials. We 
were able to obtain a completely bosonized effective action for this non-local
field theory, and we showed that it contains the TL model as a special case.

The main purpose of the present article is to extend the path-integral 
approach to non-local bosonization proposed in \cite{NRT}, to the case in
which an interaction between the electrons and a finite density of fermionic
impurities is included in the action. This generalization of the non-local
bosonization procedure provides a new way to examine the low-energy physics
of the TL model in the presence of localized impurities, that could allow to
make contact with recent very interesting studies \cite{Sch'},\cite{GiS} on
the response of a LL to localized perturbations.

We describe the impurities following the work of Andrei \cite{Andrei}, 
who introduced a new fermionic field with vanishing kinetic energy to represent 
a finite density of impurities, arbitrarily (not randomly) 
situated. This treatment has been previously employed, for example, in the 
path-integral bosonization of the Kondo problem \cite{FS}.

We introduce a non-local diagonal potential matrix binding impurities 
and electrons through their corresponding fermionic currents. This procedure 
allows to treat a wide range of possible interactions, depending on 
the precise functional form of the potential matrices. 
The complete coupling term includes interactions between charge, current, 
spin and spin-current densities. Our functional approach enables us to obtain 
an effective action governing the dynamics of the collective modes, providing 
then a practical framework to face a non-perturbative analysis of bosonic 
degrees of freedom in the presence of impurities.
We start from the partition function
\begin{equation}
Z = \int D\bar{\Psi}~D\Psi~D\bar d~Dd~e^{-S},
\label{a}
\end{equation}
where the action $S$ can be splitted as
\begin{equation}
S = S_0 + S_{int},
\end{equation}
with
\begin{equation}
S_0  =  \int d^2x~ [\bar{\Psi} i \raise.15ex\hbox
{$/$}\kern-.57em\hbox{$\partial$} \Psi + d^{\dagger} i \partial_t d] 
\end{equation}
and
\begin{equation}
S_{int} = - \int d^2x~d^2y~ [J^a_{\mu}(x) V^{ab}_{(\mu)}
(x,y)J^b_{\mu}(y) + J^a_{\mu}(x) U^{ab}_{(\mu)}(x,y) S^b_{\mu}
(y)],
\label{1}
\end{equation}
where the electron field $\Psi$ is written as
\[ \Psi = \left( \begin{array}{c} 
              \Psi_{1} \\
              \Psi_{2} \\
              \end{array} \right),  \]
with $\Psi_1$ ($\Psi_2$) in the fundamental representation of  
U(N), describing right (left) movers, whereas the impurity field $d$
is given by
\[ d = \left( \begin{array}{c}  
                            d_{1}  \\
                            d_{2}  \\
                            \end{array} \right). \]
Note the absence of a spatial derivative in the free piece of the 
impurity action, meaning that the corresponding kinetic energy is zero.
Concerning the electronic kinetic energy, we have set the Fermi velocity
equal to 1.\\
The interaction pieces of the action have been written in terms of 
U(N) currents $J_{\mu}^{a}$ and $S_{\mu}^{a}$, defined as
\begin{eqnarray}
J_{\mu}^{a} &=& \bar{\Psi} \gamma_{\mu} \lambda^{a} \Psi,  \nonumber\\
S_{\mu}^{a} &=& \bar d \gamma_{\mu} \lambda^{a} d,~~~ a = 0,1,...,N^2-1,
\label{,}
\end{eqnarray}
with $\lambda^0= I/2$, $\lambda^j= t^j$, $t^j$  being the SU(N)
generators normalized according to $tr (t^i t^j ) = \delta^{ij}/2$. 
$V^{ab}_{(\mu)}(x,y)$ and $U^{ab}_{(\mu)}(x,y)$ are $N^2 \times 
N^2$ matrices whose elements are symmetric bilocal arbitrary potentials 
describing the electron-electron (e-e) and the electron-impurity (e-i) 
interactions, respectively.  
Note that no sum over repeated indices is implied when a subindex $(\mu)$ 
is involved. 

In the next Section we shall show how to extend the above mentioned 
approach to non-local bosonization \cite{NRT}, originally developed for the 
impurity-free Thirring model, to the case in which e-i interactions 
are considered. We will express the partition function (\ref{a}) in terms 
of two fermionic determinants, one associated to conduction electrons and 
the other one to the fermionic impurities.

In Section 3 we solve these determinants for the non-Abelian (U(2)) 
case. This model describes the general forward-scattering problem of spin 1/2
electrons coupled to impurities, with both e-e and e-i spin-changing
interactions. We obtain an effective bosonic action including 
Wess-Zumino-Witten-type terms. Although the structure of this action is quite 
involved, it exhibits spin-charge decoupling explicitly.

In Section 4 we specialize the results of previous sections to the case 
in which the maximal Abelian subgroup of U(2) is considered. This model 
represents a many-body system of spin 1/2 particles when spin-flipping 
processes are disregarded but non-trivial interactions with fermionic 
impurities are taken into account. In Subsection 4.1 we obtain the completely 
bosonized action describing the collective modes of the system. From this 
action we get the dispersion relations of these oscillations as functions of 
e-e and e-i interaction potentials. Two-point fermionic correlations and 
electronic momentum distributions are computed in 4.2 and 4.3, respectively. 
In this last Subsection we illustrate a possible application of our work by 
discussing a "potential tuning" mechanism that leads to a restoration of the 
Fermi edge. Although we were able to show this effect only for a very special 
choice of the couplings, which evidently weakens its experimental relevancy, 
we think it deserves attention as a first step towards a reconciliation between 
the standard TL model and the FL phenomenology.

In Section 5 we summarize our results and conclusions.

We also include an Appendix that gathers the main results on fermionic 
and bosonic Jacobians associated to the changes of variables used throughout 
the paper.

\section{Partition function in terms of fermionic determinants}
\setcounter{equation}{0}

\indent Our first goal is to express the functional integral (\ref{a}) in terms 
of fermionic determinants. To this end we define new currents \\
\begin{equation}
K^a_{\mu}(x) = \int d^2y~ V^{ab}_{(\mu)}(x,y)J^b_{\mu}(y),
\label{KK}
\end{equation}
and
\begin{equation}
P^a_{\mu}(x) = \int d^2y~ U^{ab}_{(\mu)}(x,y)J^b_{\mu}(y).
\label{PP}
\end{equation} \\
The usual procedure in order to match the quartic interactions 
between fermions consists in introducing auxiliary fields. In the present case 
the action contains two terms quadratic in currents, so one needs two fields 
$A^a_{\mu}(x)$ and $B^a_{\mu}(x)$, which are incorporated in the functional
integrand in the form
\begin{eqnarray}
Z & = &\int D\bar{\Psi}~D\Psi~D\bar d~Dd~e^{-S_0 }\nonumber\\
&\int & DA^a_{\mu}~ \delta [A^a_{\mu} - K^a_{\mu}] exp [
\int d^2x~J_{\mu}^{a}A^a_{\mu}]\nonumber\\
& \int & DB^a_{\mu}~ \delta [B^a_{\mu} 
- P^a_{\mu}] exp[ \int d^2x~S_{\mu}^{a}B^a_{\mu}].
\end{eqnarray}
In order to fix the number of impurities to be $n$, we introduce a 
Lagrange multiplier $a$, enforcing the constraint $d^{\dagger}d = n$. 
It becomes then natural to redefine the free piece of the action as
\begin{equation}
S'_0 = S_0 + \int d^2x~a(d^{\dagger}d - n).
\end{equation}
On the other hand, we represent the $\delta$ functionals as integrals of 
exponentials over two new fields $C^a_{\mu}(x)$ and $D^a_{\mu}(x)$, thus 
obtaining
\begin{eqnarray}
Z & = & \int  D\bar{\Psi}~D\Psi~D\bar d~Dd~Da~DA^a_{\mu}~DB^a_{\mu}~e^{-S'_0 
(\Psi, d)}\nonumber\\ 
& &exp[ \int d^2x~J_{\mu}^{a}A^a_{\mu}+
\int d^2x~ S_{\mu}^{a}B^a_{\mu}]
\int DC^a_{\mu} exp[-\int d^2x~(A^a_{\mu} - K^a_{\mu})C^a_{\mu}] \nonumber\\
& \int & DD^a_{\mu} exp[-\int d^2x~(B^a_{\mu} - P^a_{\mu})D^a_{\mu}].
\label{k}
\end{eqnarray}
At this point one sees that the fermionic piece of the action
(the free part and the terms involving the currents J, S, K and P) can be cast
in a local form by defining the "potential transformed" fields
\begin{equation}
\bar{C}^a_{\mu}(x) = \int d^2y~ V^{ab}_{(\mu)}(x,y) C^b_{\mu}(y),
\end{equation}
\begin{equation}
\bar{D}^a_{\mu}(x) = \int d^2y~ U^{ab}_{(\mu)}(x,y) D^b_{\mu}(y),
\end{equation}
We then get
\begin{eqnarray}
Z& = &\int D\bar{\Psi}~D\Psi~D\bar d~Dd~Da~DA~DB~D\bar{C}~D\bar{D}\nonumber\\
& &exp\{-\int d^2x~[\bar{\Psi}( i \raise.15ex\hbox{$/$}\kern-.57em\hbox
{$\partial$} + (\not\!\! A + \not\!\! \bar{C} + \not\!\! \bar{D}))\Psi + \bar 
d (i \gamma_0 \partial_t + \gamma_0 a +  \not\!\! B) d +\nonumber\\
&+& C^a_{\mu}(x)A^a_{\mu}(x) + B^a_{\mu}(x)D^a_{\mu}(x) - a.n]\}.
\end{eqnarray}
This equation, in turn, suggests the following change of variables
\[
A^a_{\mu} + \bar{C}^a_{\mu} + \bar{D}^a_{\mu} = \tilde{A}^a_{\mu},
\]
\begin{equation}
A^a_{\mu} - \bar{C}^a_{\mu} + \bar{D}^a_{\mu} = \tilde{C}^a_{\mu},
\label{13'}
\end{equation}
\[
A^a_{\mu} + \bar{C}^a_{\mu} - \bar{D}^a_{\mu} = \tilde{D}^a_{\mu},
\]
giving
\begin{eqnarray}
Z = &\int & D\tilde{A}~DB~D\tilde{C}~D\tilde{D}~det(i\raise.15ex\hbox{$/$}
\kern-.57em\hbox{$\partial$} + \not\!\! \tilde{A})~ det(i \gamma_0\partial_t + 
i \gamma_0 a + \not\!\! B)\nonumber\\
& &\int Da~ \exp{\int d^2x~a.n} exp\int d^2x~d^2y~\{- \frac{a^{ab}_{(\mu)}(x,y)}
{2}[\tilde{A}^a_{\mu}(x)-\tilde{D}^a_{\mu}(x)]B^b_{\mu}(y)\}\nonumber\\
& & exp\int d^2x~d^2y~\{- \frac{b^{ab}_{(\mu)}(x,y)}{4}
[\tilde{A}^a_{\mu}(x)-\tilde{C}^a_{\mu}(x)][\tilde{D}^b_{\mu}(y)+
\tilde{C}^b_{\mu}(y)]\},
\end{eqnarray}
where we have defined the inverse potentials $a$ and $b$ through
the identities
\begin{equation}
\int V^{ac}_{(\mu)}(x,y) b^{cd}_{(\mu)}(x,z) d^2x~ = \delta^{ad}
\delta^{(2)}(y-z),
\label{aa}
\end{equation}
\begin{equation}
\int U^{ac}_{(\mu)}(x,y) a^{cd}_{(\mu)}(x,z) d^2x~ = \delta^{ad}
\delta^{(2)}(y-z),
\label{bb}
\end{equation}
In the above expression for $Z$ one sees that, by virtue of the 
change of variables (\ref{13'}), the fields $\tilde{C}$ and 
$\tilde{D}$ play no direct role in the fermionic determinants. They
are actually artefacts of our method, whereas the fields $A$ and $B$
describe the physically relevant bosonic degrees of freedom. Therefore, the
next step is to perform the integrals in $\tilde{C}$ and 
$\tilde{D}$. This can be easily done, as usual, by conveniently shifting
the fields. The field $\tilde{C}$ describes negative metric states, a 
situation already encountered in the impurity-free problem \cite{NRT}. The 
important difference is that here the ghost-field is coupled with both $A$ and 
$B$. One has then to first perform the corresponding integration in order to 
extract the dependence on the relevant fields, and only when this is done, to 
absorb the decoupled ghost partition function in the overall normalization 
constant. This treatment of the negative metric states is in agreement with the 
prescription of Klaiber \cite{Kl} which, in the operator framework, precludes
the use of an indefinite-metric Hilbert space. Taking all these considerations
into account, and setting from now on $\tilde{A}=A$, one finally gets
\begin{equation} 
Z =  \int DA~DB Da~ e^{-S'[A,B]} det ( 
i\raise.15ex\hbox{$/$}\kern-.57em\hbox{$\partial$} +
\not \!\! A ) det ( i \gamma_{0} \partial_{t} + a \gamma_0 +
\not \!\! B ).
\label{6}
\end{equation}
with
\begin{eqnarray}
S'[A,B] & = & \int d^2x a.n + \int d^2x ~ d^2y [ B_{\mu}^{a}(x) 
a_{(\mu)}^{a b}(x,y) A_{\mu}^{b}(y)  +  \nonumber \\ 
& - & B_{\mu}^{a}(x) \int d^2u ~ d^2v ~ a_{(\mu)}^{a c}(x,u) 
( b_{(\mu)}^{-1})^{c d}(u,v) a_{(\mu)}^{d b}(v,y) B_{\mu}^{b}(y)]. 
\label{7}
\end{eqnarray}
This is our first non-trivial result. We have been able to express  
$Z$ in terms of two fermionic determinants, one dependent on the original 
electronic fields coupled to $A$, and the other one defined through 
the impurity degrees of freedom interacting with $a$ and $B$. Note that 
although these fermionic determinants involve independent bosonic fields, the 
bosonic action provides an interaction term coupling them. The important point 
is that now, the non-local character of the interactions is entirely located 
in this bosonic action, whose functional form depends on the original 
couplings. 
Interestingly enough, however, one can go further quite a long way without 
specifying the potentials. This will be shown in the next Sections, where we
shall deal explicitly with the determinants.
                                            
\section{The spin-flipping (non-abelian) case}
\setcounter{equation}{0}
In the context of 2d QFT's, the bosonization procedure \cite{Stone}, 
originally developed in the operator language \cite{Low}, has been also 
fruitfully implemented in the path-integral framework \cite{GS}. In this 
Section we show how to use this last method, based on decoupling changes in 
the functional measures, in order to get an effective bosonic action for the 
general forward-scattering problem defined in the previous sections.
The technique depends non-trivially on the nature of the groups involved. 
Here we consider the case in which fermion fields are taken in the 
fundamental representation of U(2) and the vector fields take values in the
Lie algebra of U(2). Although our procedure can be easily extended to the U(N)
case, bearing in mind that we want to make contact with the spin 1/2 TL model,
we set N = 2 from the beginning.\\
\indent Let us start by considering the first determinant in (\ref{6}), 
associated with the conduction electrons. 
We decompose the field $A_{\mu}$ as
\begin{equation}
A_{\mu} =  \Lambda_{\mu} + {\cal A}_{\mu},
\end{equation}
with
\begin{equation}
\Lambda_{\mu} = \Lambda^0_{\mu} \lambda^0, 
\end{equation}
and
\begin{equation}
{\cal A}_{\mu} = {\cal A}_{\mu}^i \lambda^i,
\end{equation}
where $\lambda^0 = I/2$ and $\lambda^i$ (i = 1,2,3) are the SU(2) 
generators. The U(1) part of the vector field can be decoupled from the 
fermionic fields by performing the following change in the path-integral 
variables:
\begin{eqnarray}
\Psi (x) &=& e^{i \eta^0 (x) + \gamma_5 \Phi^0(x)}\chi \nonumber\\ 
\bar{\Psi} (x) &=& \bar{\chi} e^{-i \eta^0 (x) + \gamma_5 \Phi^0(x)}\nonumber\\ 
\Lambda_{\mu}^0(x) &=& - \epsilon_{\mu \nu} \partial_{\nu}\Phi^0 + 
\partial_{\mu} \eta^0.
\label{ab}
\end{eqnarray}
where $\Phi^0$ and $\eta^0$ are scalars.
\noindent The Jacobian associated to the change in the bosonic variables is 
trivial, whereas that related to the fermionic variables gives
\begin{equation}
J_{U(1)}^F =\exp{-\frac{1}{2\pi} \int d^2x~(\partial_{\mu}\Phi^0)^2}.
\end{equation}
We then have
\begin{equation}
\int DA_{\mu} det( i \raise.15ex\hbox{$/$}\kern-.57em\hbox{$\partial$} + 
\not\!\! A ) =  \int D\Phi^0 D\eta^0 D{\cal{A}}  
e^{-\frac{1}{2\pi}\int d^2x~(\partial_{\mu} \Phi^0)^2} det( i \raise.15ex
\hbox{$/$}\kern-.57em\hbox{$\partial$} + \not\!\!\cal{A})
\label{11a}
\end{equation}
As explained in the Appendix, the SU(2) components of the vector
field can be also decoupled by using a non-Abelian extension of equations
(\ref{ab}) in the form
\begin{eqnarray}
\Psi_1 &=& h^{-1} \chi_1,\nonumber\\
\bar{\Psi}_1 &=& \bar{\chi}_1 h,\nonumber\\
\Psi_2 &=& g^{-1} \chi_2,\nonumber\\
\bar{\Psi}_2 &=& \bar{\chi}_2 g,\nonumber\\
{\cal A}_+& =& h^{-1} \partial_+h \nonumber\\
{\cal A}_-& =& g^{-1} \partial_-g
\label{14}
\end{eqnarray}
where $g$ and $h$ are elements of SU(2) and 
$\partial_\pm = \partial_0 \pm i \partial_1$. 
In contrast to the Abelian change, now both the bosonic and fermionic 
Jacobians are non-trivial. Gathering their contributions one obtains: 
\begin{eqnarray}
\int D{\cal A}~ det ( i \raise.15ex\hbox{$/$}\kern-.57em\hbox{$\partial$} + 
\not\!\!\cal{A}) &=& det (i \raise.15ex\hbox{$/$}\kern-.57em\hbox
{$\partial$}) \int Dh~Dg~ exp \{- 5 \left( W[hg^{-1}]+\right.\nonumber\\ 
&+& \left.\alpha~ tr~\int d^2x~h^{-1}\partial_+h~ g^{-1}\partial_-g \right) \}
\end{eqnarray}
where $W$ is the Wess-Zumino-Witten action defined in the Appendix 
and $\alpha$ is a regularization parameter (for a careful study of 
regularization ambiguities see \cite{Cabra}).\\
\indent The evaluation of the determinant corresponding to the impurity field 
is more subtle, and it can be performed through different methods.
(See for example the Wilson loop approach and the geometric quantization 
described in \cite{FS}). 
In the present context we prefer to apply a technique similar to the analytic 
continuation sketched in \cite{FS}.
As a first step, by writing the scalar field $a$ in terms of a 
new bosonic field $\Phi$ as 
\begin{equation}
a = 2 i \partial_t \Phi,
\end{equation}
it is easy to decouple it from the remaining determinant
\begin{eqnarray}
det( i \gamma_{0} \partial_{t} + \gamma_0 a +  \not \!\! B ) &=& 
exp\{ \frac{1}{\pi} \int d^2x (\Phi \partial_t^2 \Phi)\}
det (i \gamma_0 \partial_t + \not\!\! B) 
\label{10}
\end{eqnarray}                        
while the change in the path-integral measure reads
\begin{equation}
Da = det (2 i \partial_t) D \Phi.
\label{10'}
\end{equation}
Note that the integral in the bosonic field $\Phi$  can be performed 
and the result absorbed in an overall normalization constant ${\cal N}$.\\
\noindent As in the previous case we write the field as the sum of two 
components, one abelian vector field $\Delta_{\mu}$ and three vector 
fields ${\cal B}_{\mu}^i$. In order to decouple the field $B$ one can write
the determinant in a factorized form
\begin{eqnarray}
det( i \gamma_{0} \partial_{t}  + \not \!\! B ) &=& 
(det i \raise.15ex\hbox{$/$}\kern-.57em\hbox{$\partial$} )^{-1} 
det \left( \begin{array}{cc}
                                0   &i \partial_+\\
                                i \partial_t + B_- & 0
                                \end{array} \right)\nonumber\\
& & det \left( \begin{array}{cc}
                                0   &i \partial_t + B_+\\
                                i \partial_- & 0
                                \end{array} \right).
\label{9}
\end{eqnarray}                        
Now each one of these determinants can be easily decoupled from 
the $\Delta_{\pm}$ ($\Delta_{\pm}= \Delta_0 \pm i \Delta_1$) fields. In the 
first determinant it is the change of variables 
\begin{equation}
\Delta_+^0 = 2i\partial_{t}\zeta^0
\end{equation}
which allows us to separate the U(1) field, whereas in the second 
one we choose
\begin{equation}
\Delta_-^0 =  -2i\partial_{t}\theta^0.
\end{equation}
The corresponding Jacobians read
\begin{equation}
exp^{-\frac{1}{2\pi} \int d^2x \zeta^0 \partial_-\partial_t\zeta^0}.   
\end{equation}
and
\begin{equation}
exp^{-\frac{1}{2\pi} \int d^2x \theta^0 \partial_t\partial_+\theta^0}   
\end{equation}
respectively.\\
\noindent The resulting determinants depend on ${\cal B}$, and can be 
treated in a similar way as the $\Lambda$ field to obtain
\begin{equation}
\int D{\cal B}_+ det \left( \begin{array}{cc}
        0   &i \partial_t + {\cal B}_+\\
 \partial_- & 0
                    \end{array} \right) = 
det \left( \begin{array}{cc}
                                0   &i \partial_t \\
                                i \partial_- & 0
                                \end{array} \right) ~ \int DU e^{-5
                                \dot{W}[U]},
\label{17'}                                
\end{equation}
with $\dot{W}[U]$ a WZW-type action given by
\begin{equation}
\dot{W}[U] = \frac{1}{8 \pi} tr \int d^2x~ \partial_t U\partial_- U^{-1}
+ \Gamma [U],
\label{18''}
\end{equation}
(please see the Appendix for the definition of the functional 
$\Gamma [U]$). The bosonic field $U$ is related to ${\cal B}_{+}$ through
\begin{equation}
{\cal B}_+ = U^{-1} \partial_+U.
\label{18'}
\end{equation}
In the same way, defining ${\cal B}_-$ as
\begin{equation}
{\cal B}_- = Q^{-1} \partial_-Q,
\label{19'}
\end{equation}
we can decouple the corresponding determinant in the form
\begin{equation}
\int D{\cal B}_- det \left( \begin{array}{cc}
        0   &i \partial_+ \\
 \partial_t + {\cal B}_- & 0
                    \end{array} \right) = 
det \left( \begin{array}{cc}
                                0   &i \partial_+ \\
                                i \partial_t & 0
                                \end{array} \right) ~ \int DQ 
                                e^{-5\tilde{W}[Q]},
\label{20'}                                
\end{equation}
where $\tilde{W}[Q]$ is another WZW-type action:
\begin{equation}
\tilde{W}[Q] = \frac{1}{8 \pi} tr \int d^2x~ \partial_- Q \partial_t Q^{-1}
+ \Gamma [Q].
\label{21'}
\end{equation}
Putting all these results together one finally has
\begin{eqnarray}
Z &=& {\cal N}  det (i \raise.15ex\hbox{$/$}\kern-.57em\hbox{$\partial$} ) 
det (2 i \partial_t) det (\gamma_0 i \partial_t) \int D\Phi^0~ D\eta^0~
D\zeta^0~D\theta^0~e^{-S^0[\Phi^0,\eta^0,\zeta^0,\theta^0]}
\nonumber\\
& &\int Dh~Dg~DU~DQ~ e^{-S_{eff}[h,g,U,Q]},
\label{22'}
\end{eqnarray}
with
\begin{eqnarray}
 S^0[\Phi^0,\eta^0,\zeta^0,\theta^0]&=&
  \frac{1}{2\pi} \int d^2x~ [(\partial_{\mu}\Phi^0)^2 
  - \zeta^0\partial_{t}\partial_-\zeta^0 - \theta^0\partial_{t}\partial_+
  \theta^0]\nonumber\\ 
&+&  \int d^2x~d^2y~ \left( 
\partial_1 \Phi^0 a_{(0)}^{00}i\partial_0(\zeta^0 - \theta^0) - 
\partial_0 \Phi^0 a_{(1)}^{00}\partial_0(\zeta^0 +\theta^0)\right.+\nonumber\\ 
&+& \partial_0 \eta^0 i a_{(0)}^{00}\partial_0(\zeta^0 - \theta^0) 
-\partial_1 \eta^0 a_{(1)}^{00}\partial_0(\zeta^0 + \theta^0) + 
\partial_0\zeta^0 c_-^{00}\partial_0\zeta^0 +\nonumber\\
&+&\left.\partial_0\theta^0 c_-^{00}\partial_0\theta^0
-2 \partial_0\zeta^0 c_+^{00}\partial_0\theta^0\right)
\label{q}
\end{eqnarray}
and
\begin{eqnarray}
S_{eff}[h,g,U,Q] &=& 5 \{ W[hg^{-1}] + \alpha~ tr \int d^2x~ 
h^{-1}\partial_+h~g^{-1}\partial_-g +\nonumber\\
&+& \dot{W}[U] +  \tilde{W}[Q]\} +\nonumber\\
&+& \frac{1}{4 } \int d^2x~d^2y~ \{ (g^{-1}\partial_+g)^a(x) (a_0 - a_1)
^{ab}(x,y)(U^{-1}\partial_+U)^b(y) +\nonumber\\  
&+& (g^{-1}\partial_+g)^a(x)(a_0 + a_1)^{ab}(x,y)
(Q^{-1}\partial_-Q)^b(y) +\nonumber\\
&+& (h^{-1}\partial_-h)^a(x)(a_0 + a_1)^{ab}(x,y)
(U^{-1}\partial_+U)^b(y) +\nonumber\\
&+& (h^{-1}\partial_-h)^a(x)(a_0 - a_1)^{ab}(x,y)
(Q^{-1}\partial_-Q)^b(y)\} -\nonumber\\
&+& \frac{1}{4} \int d^2x~d^2y~ \{ (U^{-1}\partial_+U)^a(x) 
(c_0 - c_1)^{ab}(x,y) (U^{-1}\partial_+U)^b(y) +\nonumber\\
&-& 2 (U^{-1}\partial_+U)^a(x)(c_0 + c_1)^{ab}(x,y) (Q^{-1}\partial_
-Q)^b(y) +\nonumber\\
&-& (Q^{-1}\partial_-Q)^a(x)(c_0 - c_1)^{ab}(x,y) (Q^{-1}\partial_-Q)^b(y) \}, 
\label{eff}
\end{eqnarray}
where we have defined the following combinations of potential matrix
elements:
\begin{eqnarray}
(c_0 \pm c_1)^{ab}(x,y) =\int d^2u~d^2v~&\{&a_0^{ac}(x,u) (b_0^{-1})^{cd}(u,v)
a_0^{db}(v,y)\nonumber\\ 
&\pm& a_1^{ac}(x,u) (b_1^{-1})^{cd}(u,v)a_1^{db}(v,y)\}.
\end{eqnarray}
Equations (\ref{22'}),(\ref{q}) and (\ref{eff}) constitute our second 
non-trivial result. We have obtained a completely bosonized effective action 
for the collective excitations of a TL model including both e-e and e-i 
spin-flipping interactions.
Concerning the partition function of the model, it has been written as a 
product of a free determinant and two purely bosonic functionals: 
one in terms of fields $\Phi^0, \eta^0, \zeta^0$ and $\theta^0$, and the other,
which includes the effects of non-locality, in terms of $h,g,U$ and $Q$. 
The other two determinants appearing in (\ref{22'}) do not depend on the 
bosonic degrees of freedom and can then be absorbed in a normalization 
constant. 
Nevertheless, they must be taken into 
account when studying other aspects of the model as, for example, its conformal 
properties, or when implementing a finite-temperature analysis.\\
\indent We can see from (\ref{22'}) that nonlocality arises only in the bosonic 
action for $g,h,U$ and $Q$, and it includes a variety of interactions between 
the bosonic degrees of freedom, related through the arbitrary potentials.\\
\indent Let us stress that, despite the involved form of the action, in our 
framework, the occurrence of spin and charge separation becomes evident for 
this general forward-scattering problem. Indeed, it is straightforward to 
show, for instance by direct comparison with the impurity-free case 
\cite{NRT}, that the fields $\Phi^0$and $\eta^0$ are related to 
the charge-density fluctuations of the system (please recall that they were 
introduced to parametrize $\Lambda$ ,the U(1) component of $A$, just the one 
associated to charge conservation). Thus, the factorized form of the partition 
function (\ref{22'}) is just a direct consequence of the decoupling between 
charge and spin-density modes. In other words, there is an explicit separation 
between $J^0_{\mu}$ and $J^i_{\mu}$ fluctuations. The relationship linking 
these currents to charge and spin densities for branch i (i = 1,2) particles 
can be seen from their definitions 
\begin{eqnarray}
\rho_i &=& \Psi^{\dagger}_{i \uparrow} \Psi_{i \uparrow} +
\Psi^{\dagger}_{i \downarrow} \Psi_{i \downarrow},\nonumber\\
\sigma_i &=& \Psi^{\dagger}_{i \uparrow} \Psi_{i \uparrow} -
\Psi^{\dagger}_{i \downarrow} \Psi_{i \downarrow}.
\label{16'}
\end{eqnarray}
By explicitly writing the currents of the conduction electrons 
$J^a_{\mu} (a=0,1,2,3)$ defined in (\ref{,}) we can identify
$J^0_0 = \rho_1 + \rho_2$ and $J^0_1 = i(\rho_2 - \rho_1)$ as those 
connected with charge densities, and $J^1_0 = \sigma_1 + \sigma_2$ and
$J^1_1 = i(\sigma_2 - \sigma_1)$ with spin densities. The remaining 
currents, $J_{\mu}^2$ and $J_{\mu}^3$, mix spin-down and up fermionic
components; they are associated to spin conservation in processes with spin 
flips. 

\section{Bosonization without spin-flips (the maximal abelian subgroup)}
\setcounter{equation}{0}
\indent When one considers the maximal abelian subgroup of U(N), the 
interactions between currents acquire a simple form. Although our procedure 
can be easily used for arbitrary N, we shall perform the analysis as 
in the previous Section, for the special case N = 2. In this case the model 
describes a many-body system of spin-$\frac{1}{2}$ fermions when 
spin-flipping processes are not allowed (A different functional approach to
this problem -in the impurity-free case- is given in \cite{LC}). 
Now, the potential matrices are diagonal whose elements can be written in 
terms of the g-functions defined by S\'olyom \cite{So} as
\begin{eqnarray}
V_{(0)}^{00}&=&\frac{1}{4}(g_{4 \parallel}+ g_{4 \perp} + g_{2 \parallel} 
+ g_{2 \perp}),\nonumber\\
V_{(0)}^{11}&=&\frac{1}{4}(g_{4 \parallel}- g_{4 \perp} + g_{2 \parallel} 
- g_{2 \perp}),\nonumber\\
V_{(1)}^{00}&=&\frac{1}{4}(-g_{4 \parallel}- g_{4 \perp} + g_{2 \parallel} 
+ g_{2 \perp}),\nonumber\\
V_{(1)}^{11}&=&\frac{1}{4}(-g_{4 \parallel} + g_{4 \perp} + g_{2 \parallel} 
- g_{2 \perp}).
\end{eqnarray}
It is straightforward to verify that the e-e interaction term in
(\ref{1}) contains the whole set of diagrams associated to forward scattering 
processes without spin-flips. Let us recall that the coupling constants for 
incident fermions with parallel spins are denoted by the susbscript $\parallel$ 
and that for fermions with opposite spins by the subscript $\perp$. In the $g_2$ 
processes the two branches (left and right moving particles) are coupled, while 
in the $g_4$ processes all four participating fermions belong to the same 
branch.
\noindent The Tomonaga-Luttinger model, with charge-density fluctuations only,
corresponds to $V_{(0)}^{11} = V_{(1)}^{11} = 0 $. In a completely analogous 
way we introduce the potentials that couple electron and impurity currents 
in the form
\begin{eqnarray*}
U_{(0)}^{0 0} & = & \frac{1}{4} ( h_{4\parallel} + h_{4\perp} + h_{2\parallel} 
+ h_{2\perp}), \\
U_{(0)}^{11} & = & \frac{1}{4} ( h_{4\parallel} - h_{4\perp} + h_{2\parallel} 
- h_{2\perp}) , \\
U_{(1)}^{0 0} &  = & \frac{1}{4} (- h_{4\parallel} - h_{4\perp} + h_{2\parallel} 
+ h_{2\perp}),  \\
U_{(1)}^{11} & = & \frac{1}{4} (- h_{4\parallel} + h_{4\perp} + h_{2\parallel} 
- h_{2\perp} ).
\end{eqnarray*}
This description includes both charge and spin density interactions, as well as 
spin-current interactions. A Kondo-like interaction, i.e. the coupling between 
spin densities only, corresponds to the case $U_{(0)}^{00} = U_{(1)}^{00} = 0$.

\subsection{Bosonized action and dispersion relations}

\indent In order to carry out the bosonization of the model we shall start from the 
partition function (\ref{6}) with the action given by (\ref{7}). 
The fermionic determinant involving the field $A_{\mu}$ can be readily 
computed by employing the same path-integral technique depicted in Section 3.
In the present case the decoupling of the vectorial field is achieved by
the following transformation
\begin{eqnarray}
\Psi & = & e^{ (\gamma_{5} \Phi + i \eta)} \chi,      \nonumber \\
\bar{\Psi} & = & \bar{\chi} e^{ (\gamma_{5} \Phi - i \eta)}, \nonumber\\
A_{\mu} & = & - \epsilon_{\mu \nu} \partial_{\nu} \Phi + \partial_{\mu} \eta,
\label{86}
\end{eqnarray}
with $\Phi = \Phi^{i} \lambda^{i} , \eta = \eta^{i} \lambda^{i}$ , $i=0,1$.

\noindent Due to the essentially non-Abelian character of this change of 
variables, the bosonic Jacobian is field-independent, and the fermionic 
determinant is decoupled in the usual way:
\begin{equation}
det ( i\raise.15ex\hbox{$/$}\kern-.57em\hbox{$\partial$} +
\not \!\! A ) = (det i\raise.15ex\hbox{$/$}\kern-.57em\hbox{$\partial$}) 
exp \frac{1}{2\pi} \int d^2x ~\Phi \Box  \Phi. 
\end{equation}
\hspace*{0.6cm} In order to treat the determinant associated with the 
impurity, once again it is convenient to first decouple the scalar field $a$
and then to write the determinant in a factorized form, as in (\ref{9}), and 
consider each factor separately. It is then easy to show that the changes
$B_+ = 2 i \partial_t \alpha$ in one of the determinants,and 
$B_- = -2 i \partial_t \beta$ in the other one, with $\alpha$ and $\beta$
scalars, together with the corresponding chiral rotations in the fermion 
variables, allow to extract free fermion contributions. The fermionic Jacobians 
associated to these transformations are
\begin{equation}
exp[-\frac{1}{2\pi}\int d^2x~ \alpha \partial_+ \partial_t \alpha].
\end{equation}
and
\begin{equation}
exp[-\frac{1}{2\pi}\int d^2x~ \beta \partial_- \partial_t \beta].
\end{equation}
All these changes determine a vacuum to vacuum functional depending on eight 
bosonic fields ($\Phi^i, \eta^i, \alpha^i$ and $\beta^i$), multiplied by the
free determinants coming from the Jacobians associated with the changes 
in the integration measures. In the present context these factors can be 
incorporated in the normalization constant. Of course, one must be more 
careful when considering a finite-temperature study, since in this case the 
free determinants would depend on temperature.

The action corresponding to the variables $\alpha$ and $\beta$ turns out to be 
quadratic and therefore the functional integrations over these fields can be 
readily performed. One is then left with a partition function in terms of the 
bosonic fields $\Phi^i$ and $\eta^i$ which, by comparison with the 
impurity-free case, one naturally identifies with the collective modes of
the system \cite{NRT}. The result is
\begin{equation}
 Z = \int D\Phi^i~ D\eta^i~exp - \left\{ S_{eff}^{00} + S_{eff}^{11} \right\}, 
 \label{87}
 \end{equation}
 where the actions, in Fourier space, are given by
 \newpage
 \begin{eqnarray}
 S_{eff}^{ii} & = & \frac{1}{(2 \pi)^{2}} \int d^2p 
[ \hat{\Phi}^{i}(p) A^{ii}(p) \hat{\Phi}^{i}(-p) + 
\hat{\eta}^{i}(p) B^{ii}(p) \hat{\eta}^{i}(-p)  + \nonumber \\
& + & \hat{\Phi}^{i}(p) \frac{C^{ii}(p)}{2} \hat{\eta}^{i}(-p) + 
\hat{\eta}^{i}(p) \frac{C^{ii}(p)}{2} \hat{\Phi}^{i}(-p) ], 
\label{88} 
\end{eqnarray}
where 
\begin{eqnarray}
 A(p) & = & \frac{1}{\Delta(p)}\left\{ \frac{p^{2}}{\pi} \Delta - a_{0} a_{1} 
 \frac{p_{1}^{2}}{\pi} + \frac{1}{2 \pi} (a_{0}^{2} p_{1}^{2} - a_{1}^{2} 
 p_{0}^{2}) -2 a_{0}^{2} a_{1}^{2} (\frac{p_{1}^{2}}{ b_{1}} + \frac{p_{0}^{2}}
 {b_{0}}) \right\}, \nonumber \\   
B(p) & = & \frac{1}{\Delta(p)}\left\{ \frac{p_{1}^{2}}{\pi} a_{0} a_{1} + 
\frac{1}{2 \pi} (a_{0}^{2} p_{0}^{2} - a_{1}^{2} p_{1}^{2}) - 2 a_{0}^{2} 
a_{1}^{2} ( \frac{p_{1}^{2}}{b_{0}} + \frac{p_{0}^{2}}{b_{1}}) \right\}, 
\nonumber \\
C(p) & = & \frac{1}{\Delta(p)}\left\{ \frac{ a_{0} a_{1}}{\pi} ( \frac{p_{1}^
{3}}{p_{0}} - p_{0} p_{1} ) + \frac{ p_{0} p_{1}}{\pi} (a_{0}^{2} + a_{1}^{2}) + 
4 p_{0} p_{1} a_{0}^{2} a_{1}^{2} (\frac{1}{ b_{0}} - \frac{1}{b_{1}})\right\},
\nonumber \\
\Delta (p)& =& \frac{ p_{1}^{2}}{ 4 \pi^{2} p_{0}^{2}} + 4 ( \frac{1}{ 4 \pi} -
\frac{ a_{1}^{2}}{ b_{1}}) ( \frac{1}{ 4 \pi } + \frac{a_{0}^{2}}{ b_{0}}).
\label{88a}
\end{eqnarray}
\hspace*{0.6cm} For the sake of clarity we have omitted $ii$ superindices in 
the above expressions, which are written in terms of the Fourier transforms 
of the inverse potentials defined in (\ref{aa}) and (\ref{bb}). (Note that 
$b_{\mu}(p) = V_{(\mu)}^{-1}(p)$ and $a_{\mu}(p) = U_{(\mu)}^{-1}(p)$ ).

\indent This is one of our main results. We have obtained a completely bosonized 
action for the collective modes corresponding to a system of electrons which 
interact not only between themselves, but also with fermionic localized
impurities at $T=0$.
This effective action describes the dynamics of charge density 
($\Phi^0$ and $\eta^0$) and spin density ($\Phi^1$ and $\eta^1$)fields. 
As we can see, these modes remain decoupled as in the impurity free case. 
Their dispersion relations can be obtained from the poles of the corresponding 
propagators. Alternatively, one can write the effective Lagrangian as
\begin{equation}
L_{eff}^{ii} = \frac{1}{2\pi} \left( \hat{\Phi}^i ~ \hat{\eta}^i \right)
\left( \begin{array}{cc}
A^{ii} & C^{ii}/2 \\
C^{ii}/2& B^{ii}
\end{array} \right) ~ \left( \begin{array}{c}
\hat{\Phi}^i\\
\hat{\eta}^i
\end{array}\right),
\label{15}
\end{equation}
with $A,B$ and $C$ as defined above, and solve the equation 
\begin{equation}
C^2(p) - 4 A(p) B(p) = 0.
\end{equation}
Going to real frecuencies: $p_0 = i \omega, p_1 = q$, this equation
has the following pair of relevant solutions:
\begin{equation}
\omega^{2}_{\rho}(q) = q^{2}  \frac{ 1 + \frac{2}{\pi} V_{(0)}^{00} + 
\frac{1}{2 \pi ^{2}}\{(U_{(0)}^{00})^{2} - 2 U_{(0)}^{00} U_{(1)}^{00}\} }
{1 + \frac{2}{\pi} V_{(1)}^{00} - \frac{1}{2 \pi ^{2}} 
(U_{(1)}^{00})^{2}},
 \label{141}
 \end{equation}
\begin{equation}
\omega^{2}_{\sigma}(q) = q^{2}  \frac{ 1 + \frac{2}{\pi} V_{(0)}^{11} + 
\frac{1}{2 \pi ^{2}}\{(U_{(0)}^{11})^{2} - 2 U_{(0)}^{11} U_{(1)}^{11}\} }
{1 + \frac{2}{\pi} V_{(1)}^{11} - \frac{1}{2 \pi ^{2}} 
(U_{(1)}^{11})^{2}},
 \label{15'}
 \end{equation}
The first equation gives the dispersion relation associated to 
charge-density fluctuations $(\hat{\Phi}^0,\hat{\eta}^0)$, whereas 
the second one corresponds to spin-density modes $(\hat{\Phi}^1,
\hat{\eta}^1)$. To understand this identification one can write the fermionic 
currents in terms of the charge and spin densities as we did in (\ref{16'}).\\
As a confirmation of the validity of our approach, we note that the above 
dispersion relations, involving both e-e and e-i interaction potentials, 
coincide with the well-known result for the spectrum of charge and spin 
excitations in the TL model without impurities, obtained by choosing 
$V_{(1)} = U_{(0)} = U_{(1)} =0$ and $V_{(0)} = v(q)$, in the above 
formulae (see \cite{NRT}). 

\subsection{Two point fermionic correlations}

\indent Let us now consider the fermionic 2-point function
\begin{equation}
\langle \Psi (x) \bar{\Psi} (y) \rangle = \left( \begin{array}{cc}
                                           o & G_1(x,y)\nonumber\\
                                    G_2(x,y) & 0 
                                          \end{array}  \right)
\end{equation}
where 
\begin{equation}
 G_{1(2)}(x,y) = \left( \begin{array}{cc}
                        G_{1(2)\uparrow}(x,y) & 0\nonumber\\
                                            0 & G_{1(2)\downarrow}(x,y) 
                        \end{array} \right)
\end{equation}
The subindex $1 (2)$ means that we consider electrons belonging to 
the branch $1 (2)$, and $\uparrow$ ($\downarrow$) indicates that the field 
operator carries a spin up (down) quantum number.
Let us recall that in the present case we have disregarded those processes 
with spin-flip. This is why the fermionic Green function do not have non-zero
components with mixed spin indices.\\
To be specific we consider $G_{1 \uparrow}$ (similar 
expressions are obtained for $G_{2 \uparrow}$, $G_{1 \downarrow}$ and $G_{2 
\downarrow}$ ). When the decoupling chiral change is performed, the components 
of the Green functions are factorized into fermionic and bosonic contributions 
in the form
  \begin{eqnarray}
  G_{1 \uparrow}(x,y) & = & <\Psi_{1 \uparrow} (x) \Psi^{\dag}_{1 \uparrow}(y)>
\nonumber \\
  & = & G_{1 \uparrow}^{(0)} (x,y) <e^{\left\{[ \Phi^{0}(y) 
  -\Phi^{0}(x)] + 
i[\eta^{0}(y) - \eta^{0}(x)]\right\}}>_{00}  \times  \nonumber \\ 
& \times & <e^{\left\{[ \Phi^{1}(y) -\Phi^{1}(x)] + 
i[\eta^{1}(y) - \eta^{1}(x)]\right\}}  >_{11},
\label{16}
\end{eqnarray}
where $G_{1 \uparrow}^{(0)} (x,y)$ is the free propagator, which
involves the Fermi momentum $p_F$, and is given by
\begin{equation}
G_{1 \uparrow}^{(0)} (x,y) = \frac{e^{i p_F z_1}}{2 \pi \mid z \mid^2} 
(z_0 + i z_1).
\label{17}
\end{equation}
The symbol $ <>_{ii}$ means v.e.v. with respect to the action 
(\ref{88}, \ref{88a}). Exactly as we did in the impurity-free case 
(\cite{NRT}), the bosonic factors in (\ref{16}) can be evaluated by
appropriately shifting the fields. Indeed, working in momentum-space, and 
defining the non-local operator
\begin{equation}
D(p;x,y) = e^{-ip.x} - e^{-ip.y},
\end{equation}
the functional integrations can be performed, yielding 
\begin{eqnarray}
<\Psi_{1 \uparrow} (x) \Psi^{\dag}_{1 \uparrow}(y)> = G_{1 \uparrow}^{(0)} 
(x,y)&& exp\{-\int \frac{d^2p}{(2 \pi)^2} D^2 \frac{A^{00} - B^{00} + i 
C^{00}}{4 A^{00}B^{00} - (C^{00})^2}\}\nonumber\\
& & exp\{-\int \frac{d^2p}{(2 \pi)^2} D^2 \frac{A^{11} - B^{11} + i C^{11}}
{4 A^{11}B^{11} - (C^{11})^2}\},\nonumber\\
\end{eqnarray}
with $A^{ii}, B^{ii}$ and $C^{ii}$ given by (\ref{88a}).
In order to continue the calculation one needs, of course, to specify the
couplings and perform the integrals. This means that our formula could be used 
to test the effect of different e-e and e-i potentials on the behavior of the 
fermionic propagator.

\subsection{Momentum distribution}

\indent As a final illustration of our procedure, in this Subsection we shall 
show how to compute the electronic momentum distribution, for a quite peculiar
choice of e-e and e-i potential matrix elements.\\
Let us consider the momentum distribution of electrons belonging to branch 1 
and with spin-up. This distribution is given by
\begin{equation}
N_{1 \uparrow}(q) = C(\Lambda) \int^{+\infty}_{-\infty} dz_1~ e^{-iqz_1}
\lim_{z_0 \rightarrow 0} G_{1\uparrow}(z_0, z_1).
\label{18}
\end{equation}
We shall set
\[
V_{(1)}^{00} = V_{(1)}^{11} = 0,~\\
V_{(0)}^{00} = \frac{\pi}{2} r,~\\
V_{(0)}^{11} = \frac{\pi}{2} s,
\]
which corresponds to an e-e interaction including only 
charge-density fluctuations (the usual TL model). 
Concerning the interaction between electrons and impurities, 
we shall take into account only spin-density and spin-current interactions,
\[
U_{(0)}^{00} = U_{(1)}^{00} = 0,\\
\left( U_{(0)}^{11}\right) ^2 = \left( U_{(1)}^{11}\right) ^2 = 2 \pi^2 t.
\]
Note that for repulsive electron-electron interactions one has 
$r > 0$ and $s > 0$, whereas $t > 0$ for both ferromagnetic and 
antiferromagnetic couplings.\\
Taking the limit $z_0 \rightarrow 0$ in (\ref{16}) and replacing  eqs. 
(\ref{16}) and (\ref{17}) in (\ref{18}) one gets
\begin{equation}
 N_{1 \uparrow}(q) = C(\Lambda) \int dz_{1} \frac{e^{-i(q - p_{F})z_{1}}}{z_{1}}
 e^{-\int dp_{1} \frac{ 1 - cos p_1 z_1 }{p_1}  \Gamma (p_1)},
 \label{19}
 \end{equation}
where $C(\Lambda)$ is a normalization constant depending on an ultraviolet 
cutoff $\Lambda$, and $\Gamma(p_1)$ depends on $p_1$ through the potentials 
in the form
\begin{eqnarray}
\Gamma (r,s,t) & = & \frac{( \mid 1 + r \mid^{1/2} -  1)^{2}}
{\mid 1 + r \mid^{1/2}} + \frac{( \mid 1 + s - t \mid^{1/2} - \mid 1 -t 
\mid^{1/2})^{2}}{\mid 1 + s - t \mid^{1/2} \mid 1 - t \mid^{1/2}} -\nonumber\\
 & - & \frac{ 2 t ( t - t_{c} )}{  \mid 1 + s - t \mid^{3/2} 
\mid 1 - t \mid^{1/2}}.
\label{20}
\end{eqnarray}
If we define
\begin{equation}
f(s,t) = \frac{( \mid 1 + s - t \mid^{1/2} - \mid 1 -t \mid^{1/2})^{2}}
{\mid 1 + s - t \mid^{1/2} \mid 1 - t \mid^{1/2}}, 
\label{21}
\end{equation}
\begin{equation}
h(s,t)= t \frac{ 2 t ( t -1 ) + s (s - 1)}{  \mid 1 + s - t \mid^{3/2} 
\mid 1 - t \mid^{1/2}},
\label{22}
\end{equation}
we can write eq. (\ref{20}) in the form
\begin{equation}
\Gamma (r,s,t) = f(r,0) + f(s, t) - h(s,t).
\label{23}
\end{equation}
Once again, in order to go further and make the integration in
(\ref{19}), one has to specify the functional form of $r$, $s$ and $t$. 
At this point one observes that we are in a position of discussing, through 
this simple example, an interesting aspect of the general model under 
consideration. Indeed, we can try to determine under which conditions it is 
possible to have a restoration of the Fermi edge. To this end, and as a first 
approximation, we shall consider contact interactions ($r$, $s$ and $t$ 
constants) and search for those relations between potentials giving 
$\Gamma (r,s,t) = 0$. 
In this last case one obtains the well-known normal 
Fermi-liquid (FL) behavior
\begin{equation}
N_{1 \uparrow}(q) \approx \Theta (q - p_F).
\label{24} 
\end{equation}
At this point some remarks are in order. 
In the impurity free case ($t = 0$), $\Gamma (r,s,t)$ cannot vanish for any 
value of $r$ and  $s$ other than $r=s=0$, which corresponds to the 
non-interacting Fermi gas. This result is consistent with the well-known
LL behavior of the TL model.\\
In order to have collective modes with real frecuencies ($\omega^2>0$), one 
finds two regions where the FL edge could be restored: $t > 1+s$ and $t<1$.

\noindent In eq.(\ref{21}) one can observe that $f(r,0)>0$, thus setting 
$\Gamma = 0$ yields the condition
\begin{equation}
F(s,t) = h(s,t) - f(s,t) > 0. 
\label{25}
\end{equation}
A simple numerical analysis of $F(s,t)$ shows that the above 
inequality is not fulfilled for $0 < t <1$. The electron-impurity coupling is 
not strong enough in this region as to eliminate the LL behavior. On the 
contrary, for $t > 1 + s$ equation (\ref{21}) can be always satisfied. 
Moreover, in this region, we obtain a surface in which the condition 
$\Gamma = 0$ provides the following analytical solution for $r$ in terms of 
$F(s,t)$
\begin{equation}
r = F^{2}/2 + 2 F + ( 1 + F/2) ( F^{2} + 4 F)^{1/2}.
\label{26}
\end{equation}
The above discussion can be summarized by identifying the following
three regions in the space of couplings:

\noindent Region I, given by $0 \leq t < 1$, in which one necesarilly has LL 
behavior. Region II, with $ 1 \leq t < s + 1$, in which the frecuency of the 
spin density excitations becomes imaginary; and region III, given by 
$t > s + 1$, where the FL behavior is admitted. In this region Eq(\ref{22}) 
defines a surface in the space of potentials on which FL behavior takes place. 
One particular solution belonging to this surface is obtained by choosing 
$s = 0$ in (\ref{26}), which yields 
\[ r(t) = \frac{2t(3t-2) + 2(2t-1)\sqrt{3t^2 - 2t}}{(t-1)^2}.\]
corresponding to the case in which the dispersion relation of the 
spin density excitations is given by $\omega^2 = q^2$. For $t$ large, $r$ 
approaches a minimum value $r_{min}= 6 + 4\sqrt{3}$, a feature that is shared 
with each curve $ s = constant$ on the "FL surface". \\
\indent In summary, we have shown how the e-i couplings can be tuned in order to 
have a restoration of the Fermi edge in a TL model of electrons interacting 
with fermionic impurities. Unfortunately, we could analytically work out this
mechanism only for a very peculiar choice of the e-i couplings, which 
evidently weakens its experimental relevancy. Besides, a more realistic study 
should include at least the backward-scattering processes. However, we think 
this discussion deserves attention as a first step towards a possible 
reconciliation between the standard TL model and the FL phenomenology.

\section{Summary and conclusions}

In this paper we have shown how to extend a path-integral approach to 
non-local bosonization \cite{NRT}, to the case in which, besides the fermionic
auto-interactions, an additional coupling with another set of fermion fields
is included in the initial action. Taking into account the results of 
\cite{NRT}, we identify the self-interacting fields as 1d electrons, and 
following the work of Andrei \cite{Andrei}, we interprete the new fields as a
finite density of (not randomly) localized (with zero kinetic energy) 
fermionic impurities. Concerning the scattering processes involving both,
electrons with themselves and electrons with impurities, we restrict our
analysis to the forward-scattering case.

As a necessary condition for the application of our technique, in 
Section 2 we were able to write the partition function of the system in terms 
of two fermionic determinants, one associated
to electrons and the other one to impurities, and both connected through
bosonic degrees of freedom.

In Section 3 we used the well-established functional methods for the 
evaluation of fermionic determinants \cite{GS} in order to decouple 
fermions from bosons, getting thus a bosonized action describing the dynamics 
of the collective modes, for the general problem, including spin-flipping 
arbitrary (symmetric in coordinates) potentials. As it happens in the 
impurity-free case, being the problem  a non-Abelian one (the group is U(2)), 
the analysis of the physical spectrum is not straightforward. Of course, 
in the present case the structure of the action is even more complicated since 
one not only has the usual WZW functional arising from the electronic 
bosonization, but also some other functionals coming from the interaction with 
the impurities. Some of these terms had been previously found in a 
path-integral treatment of the Kondo problem \cite{FS} which followed the same 
Andrei's scheme to represent the impurity.
Despite these drawbacks, our method has the merit of explicitly displaying
the expected decoupling between charge-density and spin degrees of freedom.
On the other hand, our effective action could be used as a starting point for
perturbative or semiclassical analysis.

In Section 4 we have considered a simpler but still interesting version of
the model which consists in restricting the generators to those expanding the
maximal Abelian subgroup of U(2). From the physical point of view this means 
that both electrons and impurities do not change their spins in the scattering
process. In this case, the action governing the dynamics of the collective
modes is quadratic and we could then obtain some analytical results. In 
particular we got exact expressions for the dispersion relations of 
charge-density (plasmons) and spin-density modes as functionals of e-e and
e-i interaction potentials. We have also computed the two-point electronic
correlation function and the corresponding momentum distribution. We want to
emphasize that our results are valid for arbitrary bilocal e-e and e-i 
potentials, i.e. in our approach one does not need to specify the couplings
in order to get closed formulae for the bosonized action and Green functions.
This is interesting because it opens the possibility of employing some of 
these results as a testing bench to check the validity of different 
potentials. Finally, as a short digression, in order to illustrate another  
possible application of our work, we have explored the behavior of the momentum
distribution for a particular choice of the potentials that corresponds to a
TL model of electrons coupled to impurities through spin-densities and 
spin-current-densities only. In this example we verified the existence of a
region in the space of couplings in which the momentum distribution recovers
the normal free-fermion form (the so called Fermi liquid behavior). However,
concerning this result we have to be very cautious, since we have disregarded
both backward-scattering and spin-flipping processes, and the resulting 
region of parameter space in which FL behavior takes place corresponds to
very large values of the couplings. Nevertheless we think that this example
could be useful as a first step towards the construction of a modified TL
model for which LL and FL behaviors could manifest as different phases within 
the same space of potentials.

This work could be followed in many directions, although we think that the main 
challenge will be to incorporate backward and umklapp processes within this 
framework. We hope to report on these issues in the close future.

\section*{Acknowledgements} 
The authors are partially supported by Fundaci\'on Antorchas (Buenos Aires, 
Argentina) under grant A-13218/1-000069.

\renewcommand{\theequation}{{\rm A}.\arabic{equation}}
\setcounter{equation}{0}
\appendix
\section*{Appendix}

\indent The decoupling method that allows to separate the bosonic part from the 
fermionic one in a Dirac determinant \cite{GS} is quite well-known
in the context of QFT's. 
Basically, the method consists in performing a change in the fermionic and 
bosonic path-integral variables, in such a way that the 
partition function becomes factorized into a bosonic factor times a fermionic 
one. For a field $A$ in the Lie algebra of a group G, parametrized in terms 
of fields $h$ and $g$ belonging to G as
\begin{eqnarray}
A_+& =& h^{-1} \partial_+h,\nonumber\\
A_-& =& g^{-1} \partial_-g,
\label{14'}
\end{eqnarray}

\noindent where $A_{\pm} = A_0 \pm iA_1, \partial_{\pm} = \partial_0 \pm i
\partial_1$, the change of variables (\ref{14'}) together with the fermionic
one
\begin{eqnarray}
\Psi_1 &=& h^{-1} \chi_1,\nonumber\\
\bar{\Psi}_1 &=& \bar{\chi}_1 h,\nonumber\\
\Psi_2 &=& g^{-1} \chi_2,\nonumber\\
\bar{\Psi}_2 &=& \bar{\chi}_2 g,\nonumber\\
\end{eqnarray}
give
\begin{eqnarray}
det( i \raise.15ex\hbox{$/$}\kern-.57em\hbox{$\partial$} + \not\!\! A ) &=& 
det i \raise.15ex\hbox{$/$}\kern-.57em\hbox{$\partial$}~ 
exp\{-\left( W[h]+W[g] \right.\nonumber\\ 
 &+&\left. (\frac{1}{4\pi} + \alpha)~ tr\int d^2x~ h^{-1}\partial_+h~
g^{-1}\partial_-g \right) \}.
\label{11}
\end{eqnarray}
In this bosonic action $W[h]$ is a  Wess-Zumino-Witten action 
term given by
\begin{equation}
W[h] = \frac{1}{8\pi}tr \int d^2x~\partial_{\mu}h^{-1}\partial^{\mu}h
+ \Gamma [h], 
\label{13}
\end{equation}
with
\begin{equation}
\Gamma [h] = \frac{1}{12\pi} tr \int_B d^3y~ \epsilon_{ijk} h^{-1}\partial^ih~
h^{-1}\partial^jh~ h^{-1}\partial^kh,
\label{13b}
\end{equation}
and $\alpha$ is an arbitrary constant related to the regularization 
ambiguities which appear when computing the fermionic Jacobian \cite{Cabra}. 
Concerning the Jacobians associated with the changes in the bosonic measure 
\cite{FNS}, they read
\begin{eqnarray}
DA_+ &=& Dh~ exp{-2CW[h]} ,\nonumber\\
DA_- &=& Dg~ exp{-2CW[g]} .
\label{11a}
\end{eqnarray}
where C is the quadratic Casimir of the group under consideration, \\
($f^{acd}f^{bcd} = \delta^{ab}C$). We can use this results, with C = N for 
G = SU(N), to write the final form of the partition function as
\begin{eqnarray}
\int DA det( i \raise.15ex\hbox{$/$}\kern-.57em\hbox{$\partial$} + \not\!\! A ) &=& det i \raise.15ex\hbox{$/$}\kern-.57em\hbox{$\partial$}~ \int Dg~Dh~ exp\{-(1+2N)\left(W[h]
+W[g]\right.\nonumber\\ 
&+& \left.(\frac{1}{4\pi} + \alpha)~ tr\int d^2x~ (h^{-1}\partial_+h~
g^{-1}\partial_-g\right)\}.
\label{11'}
\end{eqnarray}
Both determinants appearing in (\ref{9}) associated with the ${\cal B}$ field 
can be decoupled by the use of analytic continuations and the method described 
above.

Let us first consider
\begin{eqnarray}
det \left( \begin{array}{cc}
                  0   &i \partial_+\\
                  i \partial_t + {\cal B}_- & 0
                  \end{array} \right)
\end{eqnarray}
By means of the analytic continuation $\partial_- \equiv \partial_t$,
and the identification ${\cal B}_+=0$ and ${\cal B}_-= i Q^{-1} \partial_tQ$ 
one obtains
\begin{equation}
\int D{\cal B}_- det \left( \begin{array}{cc}
                  0   &i \partial_+\\
                  i \partial_t + {\cal B}_- & 0
                  \end{array} \right) = det \left( \begin{array}{cc}
                                                      0   &i \partial_+\\
                                               i \partial_t & 0
                                               \end{array} \right)  
                  \int DQ~e^{-(1 + 2 N) \tilde{W}[Q]},
\end{equation}

\noindent where $\tilde{W}[Q]$ is a WZW-type action which differs from 
(\ref{13}), because of the analytic continuation, in the first term:
\begin{equation}
\tilde{W}[Q] = \frac{1}{8 \pi} tr \int d^2x~ \partial_- Q \partial_t Q^{-1}
+ \Gamma [Q].
\end{equation}

\noindent In the same vein, using the analytic continuation $\partial_+
\equiv \partial_t$, and the identification ${\cal B}_-=0$ and ${\cal B}_+= i 
U^{-1} \partial_tU$ one obtains
\begin{equation}
\int D{\cal B}_+ det \left( \begin{array}{cc}
                   0   &i \partial_t + {\cal B}_+\\
                   i \partial_- & 0
                   \end{array} \right) = det \left( \begin{array}{cc}
                                            0   &i \partial_t  \\
                                            i \partial_- & 0
                                            \end{array} \right)~\int DU~
                                            e^{-(1 + 2 N) \dot{W}[U]},
\end{equation}

\noindent where $\dot{W}[U]$ also differs from the usual WZW action in the
surface term

\begin{equation}
\dot{W}[U] = \frac{1}{8 \pi} tr \int d^2x~ \partial_t U\partial_- U^{-1}
+ \Gamma [U].
\end{equation}

\end{document}